\documentclass[superscriptaddress, showpacs, 10pt]{revtex4}
\usepackage{mathrsfs}
\usepackage{amssymb}
\usepackage{amsmath}
\usepackage{graphicx}

\begin{document}%

\newcommand{\ket}[1]{|#1\rangle}
\newcommand{\bra}[1]{\langle#1|}
\newcommand{\inner}[2]{\langle#1|#2\rangle}
\newcommand{\lr}\longrightarrow
\newcommand{\ra}\rightarrow
\newcommand{\tr}{{\rm Tr}}
\newcommand{\sgn}{{\rm sgn}}
\newcommand{\fsp}{{\rm span}}
\newcommand{\fsup}{{\rm supp}}
\newcommand{\fdg}{{\rm diag}}

\newtheorem{thm}{Theorem}
\newtheorem{Prop}{Proposition}
\newtheorem{Coro}{Corollary}
\newtheorem{Lemma}{Lemma}
\newtheorem{Def}{Definition}

\title{Adiabatic Quantum Counting by Geometric Phase Estimation}
\author{Chi Zhang}
\email{cz2165@columbia.edu} \affiliation{Department of Computer
Science, Columbia University, New York, USA, 10027}
\author{Zhaohui Wei}
\email{weich03@mails.tsinghua.edu.cn}\affiliation{State Key
Laboratory of Intelligent Technology and Systems, Department of
Computer Science and Technology, Tsinghua University, Beijing,
China, 100084}\affiliation{Center for Quantum Technologies, National
University of Singapore, Singapore, 117542}
\author{Anargyros Papageorgiou} \email{ap@cs.columbia.edu} \affiliation{Department of Computer
Science, Columbia University, New York, USA, 10027}
\date{\today}

\begin{abstract}

We design an adiabatic quantum algorithm for the counting problem,
i.e., approximating the proportion, $\alpha$, of the marked items in
a given database. As the quantum system undergoes a designed cyclic
adiabatic evolution, it acquires a Berry phase $2\pi\alpha$. By
estimating the Berry phase, we can approximate $\alpha$, and solve
the problem. For an error bound $\epsilon$, the algorithm can solve
the problem with cost of order $(\frac{1}{\epsilon})^{3/2}$, which
is not as good as the optimal algorithm in the quantum circuit
model, but better than the classical random algorithm. Moreover,
since the Berry phase is a purely geometric feature, the result may
be robust to decoherence and resilient to certain noise.


\end{abstract}

\maketitle

\section{Introduction}

Quantum algorithms can solve certain problems significantly faster
than the known classical algorithms. However, one main obstacle to
realize the faster quantum algorithms is the decoherence induced by
the coupling environment. To overcome the obstacle, a novel quantum
computation model, quantum adiabatic computation, is a promising
candidate
\cite{adia_2,adia_5,adia_1,adia_3,adia_7,adia_new,adia_4,adia_6}.
The model is believed to enjoy inherent robustness against the
impact of decoherence in \cite{adia_3,adia_7,adia_new}. Even though
not all researchers share this view \cite{adia_4,adia_6}, adiabatic
quantum computation still attracts considerable attention. It has
been proved to be polynomially equivalent to the quantum circuit
model \cite{adia_5,adia_2}. For instance, in the adiabatic model
searching an unordered database requires time of the same order of
magnitude as Grover's algorithm \cite{grove, adia_1}, but few
adiabatic algorithms are known that have performance similar to that
of the corresponding algorithm in the quantum circuit model. In this
paper, we show an adiabatic algorithm for the counting problem. The
task of the counting problem is to approximate the proportion of
marked items in an $N$-item database, which is denoted by $\alpha$.
In classical computation, the counting problem needs $O(N)$
evaluations in the worst case setting, and
$O((\frac{1}{\epsilon})^2)$ in the randomized setting, for error
$\epsilon$. There exists a quantum algorithm solving the problem in
$O(\frac{1}{\epsilon})$ evaluations \cite{quantum_count}. Moreover,
Nayak and Wu \cite{quantum_count_opt} showed that the quantum
algorithm given in \cite{quantum_count} is optimal in the quantum
circuit model. The counting problem is also central for many
continuous problems, such as high dimensional integration, path
integration \cite{integral1,integral2,integral3} and eigenvalue
approximation \cite{eigenvalue}.

Our algorithm for the counting problem is based on the Berry phase
acquired in the adiabatic evolution. When a quantum system undergoes
a cyclic adiabatic evolution, it acquires a geometric phase, which
is known as the Berry phase \cite{Berry}. In the algorithm, we
design an adiabatic evolution such that the resulting Berry phases
encode the solution of the problem. Then we can solve the problem by
estimating the Berry phase after the adiabatic evolution. Since
Berry phases are global phases and cannot be measured directly, we
let two parts of a superposition undergo the same cyclic adiabatic
evolution in different directions. After the adiabatic evolution,
the dynamic phases of the two parts cancel out, and the Berry phases
are $\pm 2\pi\alpha$. Then, by estimating the relative phase between
the two parts of the superposition, we can estimate $\alpha$. In the
static model, the algorithm has a runtime of order
$(\frac{1}{\epsilon})^{3/2}$, which beats the optimal classical
algorithm in the randomized setting. Usually, in adiabatic
algorithms it is the final state that encodes the solution of the
problem, while in our algorithm it is the Berry phase. Since the
Berry phase is a purely geometric feature, i.e., it only depends on
the path of evolution, and is independent of details of how the
evolution is executed, the result is resilient to certain small
errors.

We begin with some preliminaries about adiabatic algorithms and the
Berry phase, which are helpful in presenting our main result.
Consider a quantum system in a state $\ket{\psi(t)}$, $(0\leq t\leq
T)$, which evolves according to the Schrodinger equation
\begin{equation}\label{SchEqu}
i\frac{d}{dt}\ket{\psi(t)} = H(t)\ket{\psi(t)},
\end{equation}
where $H(t)$ is the Hamiltonian of the system at time $t$. If the
system is initially in its ground state, and the Hamiltonian varies
slowly enough, it will remain close to the ground state of $H(t)$,
at time $t$. Let $\ket{E_0(t)}$ be the ground state of the
Hamiltonian, and $E_0(t)$ be the corresponding eigenvalue. If $H(T)
= H(0)$, then $\ket{\psi(T)}$ is close to $\ket{\psi(0)}$, with the
exception of a global phase. The phase can be divided into two
parts: the dynamic phase
\begin{equation}
\theta = -\int_{0}^{T} E_0(t)dt,
\end{equation}
and the geometric Berry phase
\begin{equation}
\gamma = i\int_{0}^{T} \inner{E_0}{\frac{d}{dt}E_0}dt.
\end{equation}
The Berry phase depends only on the path taken, not on how fast the
path is traversed. Hence, if we design a cyclic path of
Hamiltonians, the Berry phase is totally determined.

The remainder of the present paper is organized as follows. In
section \ref{evolution}, we provide the basic adiabatic evolution
used in the algorithm, which can encode the solution to a Berry
phase of a quantum system. Then we give the adiabatic algorithm for
the counting problem. In section \ref{accuracy}, we will show the
relationship between the accuracy of the algorithm and the time it
used. In the Appendix, we show the detailed derivation of the
difference of real relative phase and the Berry phase.

\section{The adiabatic algorithm for the counting problem}\label{evolution}

In this section, we will show how to encode the solution to the
Berry phases. We use a function
$f:\{0,\cdots,N-1\}\rightarrow\{0,1\}$ to denote whether an item is
marked, i.e., $f(s) = 1$, if the $s$-th item is marked; $f(s) = 0$,
otherwise. For an error bound $\epsilon$, the algorithm has $m =
\log(\frac{1}{\epsilon})$ adiabatic evolutions. In each evolution,
we use $n+1$ qubits, where $n = \log N$. The quantum state can be
divided into two systems: the control system, which has $1$ qubit,
and the computing system, which has $n$ qubits. The computing system
is in an $N$-dimensional Hilbert space, whose basis states are
denoted as $\ket{k}$, where $k=0,\cdots,N-1$. We use an equal
superposition of all basis states as the initial state in the
computing system
\begin{equation}
\ket{\psi_0} = \frac{1}{\sqrt{N}} \sum_{k=0}^{N-1}\ket{k},
\end{equation}
and use $\frac{1}{\sqrt{2}}(\ket{0}+\ket{1})$ as the initial state
in the control system. Hence, the initial state of the whole system
is
\begin{equation}\label{is}
\ket{\psi(0)} = \frac{1}{\sqrt{2}}(\ket{0}+\ket{1})\otimes
\ket{\psi_0}.
\end{equation}
In each evolution, the initial Hamiltonian of the computing system
is
\begin{equation}
H_0 = I - \ket{\psi_0}\bra{\psi_0}.
\end{equation}
Typically,in quantum algorithms, the fundamental oracle used is
\begin{equation}
\ket{x}\ket{y} \rightarrow \ket{x}\ket{y\oplus f(x)}.
\end{equation}
In our algorithm, we modify the oracle to
\begin{equation}
\ket{x}\ket{y}\ket{z} \rightarrow \ket{x}\ket{y\oplus
f(x)}\ket{z\oplus(y\cdot f(x))}.
\end{equation}
With a $2$-qubit auxiliary state
$\frac{1}{\sqrt{2}}(\ket{0}+i\ket{1})\otimes\frac{1}{\sqrt{2}}(\ket{0}-\ket{1})$,
using the oracle on the initial state, and then discarding the
auxiliary qubits, the operation of the oracle can be written as
\begin{equation}
\frac{1}{\sqrt{N}}\sum_{k=0}^{N-1}\ket{k} \rightarrow
\frac{1}{\sqrt{N}}\sum_{k=0}^{N-1}\exp(-i\frac{\pi}{2}f(k))\ket{k}.
\end{equation}
Let $\beta = 1 - \alpha$, $M = \alpha N$, and
\begin{equation}
\begin{split}
\ket{\hat{0}} &= \frac{1}{\sqrt{N-M}}\sum_{s:f(s)=0}\ket{s},\\
\ket{\hat{1}} &= \frac{1}{\sqrt{M}}\sum_{s:f(s)=1}\ket{s},
\end{split}
\end{equation}
the initial state can be rewritten as
\begin{equation}
\ket{\psi_0} = \sqrt{\beta}\ket{\hat{0}} +
\sqrt{\alpha}\ket{\hat{1}},
\end{equation}
and the states after repeatedly using the oracle are
\begin{equation}
\ket{\psi_k} = \sqrt{\beta}\ket{\hat{0}} +
(-i)^k\sqrt{\alpha}\ket{\hat{1}},
\end{equation}
for $k=1,2,3$.

In the adiabatic algorithm, we define four Hamiltonian oracles as
\begin{equation}\label{H}
H_k = I - \ket{\psi_k}\bra{\psi_k},
\end{equation}
for $k=0,1,2,3$. Then consider a linear interpolation between the
four oracles, that is
\begin{equation}
H(t) = \sum_{k=0}^3 s_k(t)H_k,
\end{equation}
where $\sum s_k(t) = 1$, for any $0\leq t\leq T$.

In the $j$-th evolution, we choose
\begin{equation}\label{S}
\begin{split}
s_0 &= \frac{1}{2}(1 + \cos \theta_j),
s_1 = \frac{1}{2}\sin \theta_j,\\
s_2 &= \frac{1}{2}(1 - \cos \theta_j), s_3 = -\frac{1}{2}\sin
\theta_j,
\end{split}
\end{equation}
where $\theta_j$ is a function from $[0,T]$ to $[0,2^j\pi]$,
satisfying $\theta_j(0) = 0$ and $\theta_j(T)=2^j\pi$. Since the
Berry phase only depends on the path of the evolution of the
Hamiltonian, the choice of the function $\theta(t)$ does not affect
our result. Combining Eq.(\ref{H}) and Eq.(\ref{S}), the Hamiltonian
of the $j$-th evolution is
\begin{equation}\label{Ham}
\begin{split}
H(\theta_j) =& I - \frac{1}{2}(1 + \cos \theta_j)
\ket{\psi_0}\bra{\psi_0} - \frac{1}{2}\sin \theta_j
\ket{\psi_1}\bra{\psi_1}\\ &- \frac{1}{2}(1 - \cos
\theta_j)\ket{\psi_2}\bra{\psi_2} + \frac{1}{2}\sin \theta_j
\ket{\psi_3}\bra{\psi_3}
\\=& I - \ket{\psi(\theta_j)}\bra{\psi(\theta_j)},
\end{split}
\end{equation}
where
\begin{equation}
\ket{\psi(\theta_j)} = \sqrt{\beta}\ket{\hat{0}} +
e^{-i\theta_j}\sqrt{\alpha}\ket{\hat{1}},
\end{equation}
and $\theta_j \in [0,2^j\pi]$. Clearly, the ground state of
$H(\theta_j)$ is $\ket{\psi(\theta_j)}$. We set the Hamiltonian of
the whole system in the $j$-th adiabatic evolution as
\begin{equation}\label{h}
\ket{0}\bra{0}\otimes H(\theta_j)+\ket{1}\bra{1}\otimes
H(-\theta_j).
\end{equation}
Then the Berry phases of the computing system after the evolution
are $\gamma_j$ and $-\gamma_j$, where
\begin{equation}\label{gam}
\gamma_j = i\int_{0}^{2^j\pi} \inner{\psi}{\frac{d}{d\theta}\psi}
d\theta = \int_{0}^{2^j\pi} \alpha d\theta = 2^j\pi\alpha.
\end{equation}
Since the dynamic phase is the same in both parts, the final state
is
\begin{equation}
\ket{\psi_j(T)} =
\frac{1}{\sqrt{2}}(e^{i\gamma_j}\ket{0}+e^{-i\gamma_j}\ket{1})
\otimes \ket{\psi_0}.
\end{equation}
Hence, at the end of the evolution, the relative phase of the first
qubit is $\Gamma_j = 2\gamma_j = 2\pi(2^j \alpha)$. In this way, we
successfully encode the solution to the relative phases of a set of
quantum states.

In the algorithm, we do not use the phase estimation procedure in
\cite{Phase estimation} to estimate $\alpha$ from the Berry phases,
in order to avoid unnecessary entanglement. We use Kitaev's
equivalent procedure instead \cite{Phase estimation2}. As described
above, in the $j$-th adiabatic evolution of the algorithm, we
prepare a quantum state whose relative phase is $2\pi(2^j\alpha)$,
for $j=1,\cdots,m$. A measurement for the first qubit in $\ket{+} =
\frac{1}{\sqrt{2}}(\ket{0}+\ket{1})$ and $\ket{-} =
\frac{1}{\sqrt{2}}(\ket{0} - \ket{1})$ basis gives the result
$\ket{+}$ with probability
\begin{equation}
p = cos^2(\gamma) = cos^2(2^j\pi\alpha).
\end{equation}
Hence, if we apply the process several times, we can approximate the
probability. More precisely, let $q = r/R$ be the ratio between the
number of $r$ of results $\ket{+}$ and the number $R$ of
measurements. Then Chernoff's bound
\begin{equation}
Prob(|p-q| \geq \delta) \leq 2e^{-2\delta^2 R}
\end{equation}
shows that for a fixed $\delta$, the error is smaller than
$\epsilon$ for only $O(\log (1/\epsilon))$ number of measurements.
In the application, we only need an error that is smaller than
$\pi/8$. Then, we obtain an estimation of $2^j\alpha$, modulo $1$,
with error $1/16$. Let
\begin{equation}
\alpha = \sum_{j=1}^{\infty} 2^{-j}\alpha_j,
\end{equation}
for $\alpha_j \in \{0,1\}$, and $\alpha_1 = 0$ since $\alpha < 1/2$.
We also use $\overline{.\alpha_1\cdots\alpha_p}$ to denote the
binary fraction $\sum_{j-1}^{p} 2^{-j}\alpha_j$. For $j=1,\cdots,
m$, we replace the known approximate value of $2^j\alpha$ by
$\eta_j$, the closet number from the set $\{0,1/8,2/8,\cdots,7/8\}$.
Hence, we have
\begin{equation}
|2^j\alpha - \eta_j|_1 < 1/16 + 1/16 = 1/8.
\end{equation}
Since if $|y-2\alpha|_1 < \delta < 1/2$, then $|y'_0-\alpha|_1 <
\delta/2$ or $|y'_1-\alpha|_1 < \delta/2$, where $y'_1,y'_2$ are the
solutions to the equation $2y'\equiv y (\text{mod} 1)$, we can start
from $2^{m}\alpha$ and increase the precision in the following way:
Set $\eta_m = \overline{.\alpha_m\alpha_{m+1}\alpha_{m+2}} = \eta_m$
and proceed by iteration:
\begin{equation}
\alpha_j = \begin{cases} 0 &\text{if}
|\overline{.0\alpha_{j+1}\alpha_{j+2}}-\eta_j|_1 < 1/4,\\
1 &\text{if} |\overline{.1\alpha_{j+1}\alpha_{j+2}}-\eta_j|_1 < 1/4,
\end{cases}
\end{equation}
for $j=m-1,\cdots,1$. By a simple induction,
$\overline{.\alpha_1\alpha_2\cdots\alpha_m}$ can estimate $\alpha$
with error less than $2^{-m} = \epsilon$.

\section{Running Time of the Adiabatic Algorithm}\label{accuracy}

In this section, we consider the accuracy of the evolutions and the
running time of the algorithm. It is easy to see that under the
Hamiltonian given in Eq.(\ref{Ham}) the actual state in the
computing system, $|\varphi(\theta)\rangle$, always stays in the
subspace spanned by $\{\ket{\hat{0}},\ket{\hat{1}}\}$. Then
$H(\theta)$ can be rewritten as
\begin{equation}
H(\theta) =
\begin{pmatrix}
\alpha &-\sqrt{\alpha\beta}e^{i\theta}\\
-\sqrt{\alpha\beta}e^{-i\theta} &\beta
\end{pmatrix}
\end{equation}
in the subspace. Assume $\omega = \frac{d\theta}{dt}$ is constant,
and $\omega\ll 1$. Let $\ket{\varphi(t)}$ be the $t$-time state in
the system which is initially in $\ket{\psi_0}$, and evolving under
$H(\omega t)$. By solving the Schrodinger equation (\ref{SchEqu}),
we attain
\begin{equation}\label{ex1}
\ket{\varphi(t)} = e^{-i\frac{1}{2}t}(Ae^{i\omega_1 t} +
Be^{i\omega_2 t})\ket{0} + e^{-i\frac{1}{2}t}(Ce^{-i\omega_1 t} +
De^{-i\omega_2 t})\ket{1},
\end{equation}
where
\begin{equation}
\omega_{1,2} = \frac{\omega \pm
\sqrt{(1-\omega)^2+4\alpha\omega}}{2},
\end{equation}
and
\begin{equation}\label{ex2}
\begin{split}
A &=
\frac{(1-\omega)^2-\alpha(1-3\omega)+(\beta-\omega)E}{(1-\omega)^2 +
4\alpha\omega +
(\beta-\alpha-\omega)E}\sqrt{\beta},\\
B &= \frac{\alpha(1+\omega)-\alpha E}{(1-\omega)^2 + 4\alpha\omega +
(\beta-\alpha-\omega)E}\sqrt{\beta},\\
C &=
\frac{(1+\omega)^2-\beta(1+3\omega)-(\alpha+\omega)E}{(1-\omega)^2 +
4\alpha\omega +
(\beta-\alpha-\omega)E}\sqrt{\alpha},\\
D &=\frac{\beta(1-\omega)+\beta E}{(1-\omega)^2 + 4\alpha\omega +
(\beta-\alpha-\omega)E}\sqrt{\alpha},
\end{split}
\end{equation}
where $E = \sqrt{(1-\omega)^2+4\alpha\omega}$, see the Appendix. On
the other hand, denote the quantum state evolving under the
Hamiltonian $H(-\theta)=H(-\omega t)$ by $\ket{\varphi'(t)}$, which
can be obtained from $\ket{\varphi}$ by exchanging all $\omega$ by
$-\omega$.

Then, the final state of the $j$-th evolution is
\begin{equation}\label{real1}
\ket{\psi_j(T)} =
\frac{1}{\sqrt{2}}(\ket{0}\ket{\varphi(T)}+\ket{1}\ket{\varphi'(T)}),
\end{equation}
where $T = 2^j\pi/\omega$, which is the time of the $j$-th
evolution, for $j =1,\cdots,m$. As indicated before, we will measure
the relative phase of the first qubit, and use it as an
approximation of $2\pi(2^j\alpha)$. Let $\ket{\varphi_\bot}$ be a
state in the span space of $\ket{\hat{0}}$ and $\ket{\hat{1}}$,
which is orthogonal to $\ket{\varphi}$, from Eq.(\ref{real1}),
\begin{equation}
\ket{\psi_j(T)} =
\frac{1}{\sqrt{2}}(\ket{0}+\inner{\varphi(T)}{\varphi'(T)}\ket{1})\ket{\varphi}\\
+
\frac{1}{\sqrt{2}}(\inner{\varphi_{\bot}(T)}{\varphi'(T)})\ket{1}\ket{\varphi_{\bot}(T)}.
\end{equation}
Hence, with probability
\begin{equation}
p_s = \frac{1+|\inner{\varphi(T)}{\varphi'(T)}|^2}{2},
\end{equation}
the relative phase will be the argument of $
\inner{\varphi(T)}{\varphi'(T)}$. It can be checked that
\begin{equation}
\inner{\varphi(T)}{\varphi'(T)}\\ =
(AA'+DD')e^{i\mu_1}+(BB'+CC')e^{-i\mu_1} +(AB'+C'D) e^{i\mu_2} +
(A'B+CD') e^{-i\mu_2},
\end{equation}
where $A',B',C',D'$ are derived from $A,B,C,D$ by exchanging
$\omega$ by $-\omega$, and
\begin{equation}
\begin{split}
\mu_1
&=\frac{1}{2}(\sqrt{(1-\omega)^2+4\alpha\omega}-\sqrt{(1+\omega)^2-4\alpha\omega})T,\\
\mu_2 &=\frac{1}{2}(\sqrt{(1-\omega)^2+4\alpha\omega}+\sqrt{(1+
\omega)^2-4\alpha\omega})T
\end{split}
\end{equation}
From the assumption $\omega\ll 1$, we have
\begin{equation}
\begin{split}
AA' + DD'
&= 1 - 3\alpha\beta\omega^2 + O(\omega^3),\\
BB'+ CC'
&=-\alpha\beta\omega^2 + O(\omega^3),\\
AB' + C'D
&= 2\alpha\beta\omega^2 + O(\omega^3),\\
A'B + CD' &=2\alpha\beta\omega^2 + O(\omega^3).
\end{split}
\end{equation}
Hence,
\begin{equation}\label{inequ0}
p_s = \frac{1+|\inner{\varphi(T)}{\varphi'(T)}|^2}{2} \geq 1 -
8\alpha\beta\omega^2,
\end{equation}
and
\begin{equation}\label{inequ1}
|arg(\inner{\varphi(T)}{\varphi'(T)}) - \mu_1| \leq
8\alpha\beta\omega^2.
\end{equation}
Moreover,
\begin{equation}\label{inequ2}
\begin{split}
\mu_1 &= 2\alpha\omega T + 2\alpha\beta(\beta-\alpha)\omega^3 T
 + O(\omega^5 T)\\
&= 2\pi(2^j\alpha)(1 + \beta(\beta-\alpha)\omega^2 + O(\omega^3)),
\end{split}
\end{equation}
in the $j$-th evolution. Hence, the difference between the expected
relative phase and $2\pi(2^j\alpha)$ is
\begin{equation}\label{error}
\Delta \leq 2\pi(1-p_s) +
|arg(\inner{\varphi_{+}(T)}{\varphi_{-}(T)}) - \mu_1| + |\mu_1 -
2\pi(2^j\alpha)| = O(2^j \omega^2).
\end{equation}

For the estimation of $2^j\alpha$ modulo $1$ with error less than
$1/16$, it is enough to make $\Delta < 2\pi/32$, and then estimate
the relative phase to its expected value within error $2\pi/32$. To
satisfy the first condition, in the $j$-th evolution we set
\begin{equation}
\omega = \omega_j = O(\frac{1}{\sqrt{2^{j}}})
\end{equation}
in Eq. \ref{error}. When estimating the relative phase, we can boost
the success probability using repetitions. In our case, $O(m-j)$
repetitions yield overall success probability greater than $1/2$.
Since the time of the $j$-th evolution is $\frac{2^j\pi}{\omega_j}$,
the total running time is
\begin{equation}\label{time}
T_{total} = \sum_{j=1}^{m} O(\frac{2^j}{\omega_j}(m-j)) =
O((2^m)^{3/2}) = O((\frac{1}{\epsilon})^{3/2}).
\end{equation}

Therefore, the adiabatic algorithm has an
$O((\frac{1}{\epsilon})^{3/2})$ running time. As we know the optimal
quantum algorithm has a running time $O(\frac{1}{\epsilon})$, our
adiabatic algorithm is not as good as the quantum algorithm given in
the circuit model. However, it is better than the best classical
algorithm, which needs $O((\frac{1}{\epsilon})^2)$ running time.

\section{Conclusion}

In conclusion, we have proposed an adiabatic algorithm for the
quantum counting problem. The key idea of the algorithm is to
construct Berry phases which equal $2\pi(2^j\alpha)$, for
$j=1,\cdots,m$, where $\alpha$ is the proportion of marked items in
the database, $m = \log(\frac{1}{\epsilon})$. The algorithm has a
running time of $O((\frac{1}{\epsilon})^{3/2})$, which beats the
classical random algorithm.

There are some special characteristics of our algorithm. Firstly,
the solution of the problem is encoded in the phase difference of
the final state, rather than the ground state of the final
Hamiltonian. The final information is achieved by measuring relative
phase, rather than the usual quantum measurement. Next, different
from usual adiabatic algorithms, such as in \cite{adia_1}, we use
more than one Hamiltonian as oracles to construct the evolution
path. The idea of designing adiabatic algorithms based on geometric
Berry phases could also be applied to solve other problems, such as
searching an unordered database \cite{future}.

We are grateful to Joseph F. Traub, Henryk Wozniakowski, Columbia
University, Luming Duan, Yongjian Han, University of Michigan, and
Mingsheng Ying, Tsinghua University, for their very helpful
discussions and comments.

\section{Appendix}

In the Appendix, we will show the details of how to derive the
quantum state under the Hamiltonian $H(\theta) = H(\omega t)$. The
quantum system is initially in a state $\ket{\psi} =
\begin{pmatrix}\sqrt{\beta},\sqrt{\alpha}\end{pmatrix}$, then
evolves under the Hamiltonian
\begin{equation}
H(\theta) =
\begin{pmatrix}
\alpha &-\sqrt{\alpha\beta}e^{i\theta}\\
-\sqrt{\alpha\beta}e^{-i\theta} &\beta
\end{pmatrix}.
\end{equation}
Let $\ket{\varphi(t)}$ be the $t$-time quantum state in the system.
Then the Schrodinger equation (\ref{SchEqu}) turns out to be,
\begin{equation}
i\frac{d}{dt}\ket{\varphi} = H \ket{\varphi} =
\begin{pmatrix}
\alpha &-\sqrt{\alpha\beta}e^{i\omega t}\\
-\sqrt{\alpha\beta}e^{-i\omega t} &\beta
\end{pmatrix}\ket{\varphi}.
\end{equation}
Let $\ket{\varphi} = \begin{pmatrix} x\\y
\end{pmatrix}$, then
\begin{equation}\label{se1}
i\frac{d}{dt} x = \alpha x - \sqrt{\alpha\beta}e^{i\omega t}y,
\end{equation}
\begin{equation}\label{se2}
i\frac{d}{dt} y = -\sqrt{\alpha\beta}e^{-i\omega t}x + \beta y.
\end{equation}
From Eq.\ref{se1},
\begin{equation}\label{drie}
\begin{split}
y &= \sqrt{\frac{\alpha}{\beta}}e^{-i\omega t}x -
i\frac{1}{\sqrt{\alpha\beta}}e^{-i\omega t}\frac{dx}{dt},\\
\frac{dy}{dt} &= \frac{1}{\sqrt{\alpha\beta}}e^{-i\omega
t}[-i\alpha\omega x +
(\alpha-\omega)\frac{dx}{dt}-i\frac{d^2x}{dt^2}].
\end{split}
\end{equation}
Substituting Eq.\ref{drie} into Eq.\ref{se2},
\begin{equation}\label{f1}
\frac{d^2x}{dt^2}+i(1-\omega)\frac{dx}{dt}+ \alpha\omega x = 0.
\end{equation}
Similarly, we can derive
\begin{equation}\label{f2}
\frac{d^2y}{dt^2}+i(1+\omega)\frac{dy}{dt}- \beta\omega y = 0.
\end{equation}
The roots of Eq.\ref{f1} and Eq.\ref{f2} are $-\frac{i}{2}\pm
i\omega_{1,2}$ separately, where
\begin{equation}
\omega_{1,2} = \frac{\omega \pm
\sqrt{(1-\omega)^2+4\alpha\omega}}{2}.
\end{equation}
Then
\begin{equation}\label{sol}
\begin{split}
x &= e^{-i\frac{1}{2}t}(Ae^{i\omega_1 t} + Be^{i\omega_2 t}),\\
y &= e^{-i\frac{1}{2}t}(Ce^{-i\omega_1 t} + De^{-i\omega_2 t}).
\end{split}
\end{equation}
From the initial state $\ket{\psi} =
\sqrt{\beta}\ket{\hat{0}}+\sqrt{\alpha}\ket{\hat{1}}$, we have
\begin{equation}
A + B = \sqrt{\beta}; C+D = \sqrt{\alpha}.
\end{equation}
Substituting Eq.\ref{sol} into Eq.\ref{se1}, it can be derived that
\begin{equation}
\frac{D}{A} = -\frac{B}{C} = \lambda =
\frac{2\sqrt{\alpha\beta}}{(\beta-\alpha-\omega)+\sqrt{(1-\omega)^2+4\alpha\omega}}.
\end{equation}
Then, we attain that
\begin{equation}\label{ex2}
\begin{split}
A &= \frac{\lambda\sqrt{\alpha}+\sqrt{\beta}}{1+\lambda^2} =
\frac{(1-\omega)^2-\alpha(1-3\omega)+(\beta-\omega)E}{(1-\omega)^2 +
4\alpha\omega +
(\beta-\alpha-\omega)E}\sqrt{\beta},\\
B &= \lambda\frac{\lambda\sqrt{\beta}-\sqrt{\alpha}}{1+\lambda^2} =
\frac{\alpha(1+\omega)-\alpha E}{(1-\omega)^2 + 4\alpha\omega +
(\beta-\alpha-\omega)E}\sqrt{\beta},\\
C &= \frac{\sqrt{\alpha}-\lambda\sqrt{\beta}}{1+\lambda^2} =
\frac{(1+\omega)^2-\beta(1+3\omega)-(\alpha+\omega)E}{(1-\omega)^2 +
4\alpha\omega +
(\beta-\alpha-\omega)E}\sqrt{\alpha},\\
D &= \lambda\frac{\lambda\sqrt{\alpha}+\sqrt{\beta}}{1+\lambda^2}
=\frac{\beta(1-\omega)+\beta E}{(1-\omega)^2 + 4\alpha\omega +
(\beta-\alpha-\omega)E}\sqrt{\alpha},
\end{split}
\end{equation}
where $E= \sqrt{(1-\omega)^2+4\alpha\omega}$. On the other hand,
denote the quantum state evolving under the Hamiltonian
$H(-\theta)=H(-\omega t)$ by $\ket{\varphi'(t)}$, and it can be
derived from $\ket{\varphi}$ by exchanging all $\omega$ by
$-\omega$.

Then, the final state of the $j$-th evolution is
\begin{equation}\label{real}
\ket{\psi_j(T)} =
\frac{1}{\sqrt{2}}(\ket{0}\ket{\varphi(T)}+\ket{1}\ket{\varphi'(T)}),
\end{equation}
where $T = 2^j\pi/\omega$, for $j =1,\cdots,m$. Let
$\ket{\varphi_\bot}$ be a state in the span space of $\ket{\hat{0}}$
and $\ket{\hat{1}}$, which is orthogonal to $\ket{\varphi}$, from
Eq.(\ref{real}),
\begin{equation}
\ket{\psi_j(T)} =
\frac{1}{\sqrt{2}}(\ket{0}+\inner{\varphi(T)}{\varphi'(T)}\ket{1})\ket{\varphi}
+
\frac{1}{\sqrt{2}}\inner{\varphi_{\bot}(T)}{\varphi'(T)}\ket{1}\ket{\varphi_{\bot}(T)}.
\end{equation}
Hence, with probability
\begin{equation}
p_s = \frac{1+|\inner{\varphi(T)}{\varphi'(T)}|^2}{2},
\end{equation}
the relative phase will be the argument of $
\inner{\varphi(T)}{\varphi'(T)}$. Since
\begin{equation}
\inner{\varphi(T)}{\varphi'(T)}\\ =
(AA'+DD')e^{i\mu_1}+(BB'+CC')e^{-i\mu_1} +(AB'+C'D) e^{i\mu_2} +
(A'B+CD') e^{-i\mu_2},
\end{equation}
where $A',B',C',D'$ are derived from $A,B,C,D$ by exchanging
$\omega$ by $-\omega$, and
\begin{equation}
\begin{split}
\mu_1
&=\frac{1}{2}(\sqrt{(1-\omega)^2+4\alpha\omega}-\sqrt{(1+\omega)^2-4\alpha\omega})T,\\
\mu_2 &=\frac{1}{2}(\sqrt{(1-\omega)^2+4\alpha\omega}+\sqrt{(1+
\omega)^2-4\alpha\omega})T
\end{split}
\end{equation}
From the assumption $\omega\ll 1$, the relative phase can be
estimated by $ \inner{\varphi(T)}{\varphi'(T)}$, from $AA'+DD'$,
$BB'+CC'$, $AB'+C'D$ and $A'B+CD'$.

Since the above $4$ terms share the same denominator, we can first
calculate the denominator $F$, then calculate their numerators
separately. The denominator is
\begin{equation}
\begin{split}
F &=
[(1-\omega)^2+4\alpha\omega+(1-2\alpha-\omega)\sqrt{(1-\omega)^2+4\alpha\omega}]\dot[(1+\omega)^2-4\alpha\omega+(1-2\alpha+\omega)\sqrt{(1+\omega)^2-4\alpha\omega}]\\
&= [(1-\omega)^2+4\alpha\omega]\cdot[(1+\omega)^2-4\alpha\omega]+
[(1+\omega)^2-4\alpha\omega](1-2\alpha-\omega)(1-(1-2\alpha)\omega+2\alpha\beta\omega^2)\\
&+
[(1-\omega)^2+4\alpha\omega](1-2\alpha+\omega)(1+(1-2\alpha)\omega+2\alpha\beta\omega^2)\\
&+[(1-2\alpha)^2-\omega^2](1-(1-2\alpha)\omega+2\alpha\beta\omega^2)(1+(1-2\alpha)\omega+2\alpha\beta\omega^2)+O(\omega^3)\\
&=1-2(1-8\alpha\beta)\omega^2 +
2(1-2\alpha)[1-2(1-5\alpha\beta)\omega^2]+(1-2\alpha)^2-\omega^2-(1-2\alpha)^2(1-8\alpha\beta)\omega^2
+
O(\omega^3)\\
&= 4\beta^2(1-2\beta(1-4\alpha)\omega^2) + O(\omega^3).
\end{split}
\end{equation}
Then
\begin{equation}
\begin{split}
&(AA' + DD')F \\&=
\beta[(1-\omega)^2-\alpha(1-3\omega)+(1-\alpha-\omega)\sqrt{(1-\omega)^2+4\alpha\omega}]\cdot[(1+\omega)^2-\alpha(1+3\omega)+(1-\alpha+\omega)\sqrt{(1+\omega)^2-4\alpha\omega}]\\
&+\alpha\beta^2[(1-\omega)+\sqrt{(1-\omega)^2+4\alpha\omega}]\cdot[(1+\omega)+\sqrt{(1+\omega)^2-4\alpha\omega}]\\
&=
\beta[(1-\omega^2)^2-2\alpha(1-5\omega^2)+\alpha^2(1-9\omega^2)+\alpha\beta(1-\omega^2)]+\beta[\beta+\beta(2\beta-1)\omega+(2\alpha-1)\omega^2](1-(1-2\alpha)\omega+2\alpha\beta\omega^2)\\
&+\beta[\beta-\beta(2\beta-1)\omega+(2\alpha-1)\omega^2](1+(1-2\alpha)\omega+2\alpha\beta\omega^2)\\
&+\beta(\beta-\omega^2)(2\alpha-1)\omega^2](1-(1-2\alpha)\omega+2\alpha\beta\omega^2)(1+(1-2\alpha)\omega+2\alpha\beta\omega^2)+O(\omega^3)\\
&=4\beta^2(1-\beta(2-5\alpha)\omega^2)+O(\omega^3).
\end{split}
\end{equation}
Hence,
\begin{equation}
AA' + DD' =
\frac{4\beta^2(1-\beta(2-5\alpha)\omega^2)+O(\omega^3)}{4\beta^2(1-2\beta(1-4\alpha)\omega^2)
+ O(\omega^3)} = 1 - 3\alpha\beta\omega^2 + O(\omega^3).
\end{equation}
\begin{equation}
\begin{split}
&(BB'+ CC')F\\
&=
\alpha^2\beta(1+\omega-\sqrt{(1-\omega)^2+4\alpha\omega})(1-\omega-\sqrt{(1+\omega)^2-4\alpha\omega})\\
&+\alpha[(1+\omega)^2-\beta(1+3\omega)-(\alpha+\omega)\sqrt{(1-\omega)^2+4\alpha\omega}]\cdot[(1-\omega)^2-\beta(1-3\omega)-(\alpha-\omega)\sqrt{(1+\omega)^2-4\alpha\omega}]\\
&= \alpha[\alpha+(9\alpha\beta+\beta^2-2)\omega^2]+
[\alpha+(1-2\alpha)\alpha\omega+(1-2\alpha)\omega^2](1-(1-2\alpha)\omega+2\alpha\beta\omega^2)\\
&+[\alpha-(1-2\alpha)\alpha\omega+(1-2\alpha)\omega^2](1+(1-2\alpha)\omega+2\alpha\beta\omega^2)\\
&+\alpha(\alpha-\omega^2)(1-(1-2\alpha)\omega+2\alpha\beta\omega^2)\cdot(1+(1-2\alpha)\omega+2\alpha\beta\omega^2)+O(\omega^3)\\
&= -4\alpha\beta^3+O(\omega^3).
\end{split}
\end{equation}
Hence,
\begin{equation}
BB'+CC' =
\frac{-4\alpha\beta^3+O(\omega^3)}{4\beta^2(1-2\beta(1-4\alpha)\omega^2)+
O(\omega^3)} = -\alpha\beta\omega^2 + O(\omega^3).
\end{equation}
\begin{equation}
\begin{split}
&(AB'+C'D)F\\ &=
\alpha\beta[(1-\omega)^2-\alpha(1-3\omega)+(\beta-\omega)\sqrt{(1-\omega)^2+4\alpha\omega}]\cdot(1-\omega-\sqrt{(1+\omega)^2-4\alpha\omega})\\
&+\alpha\beta[(1-\omega)^2-\beta(1-3\omega)+(\alpha-\omega)\sqrt{(1+\omega)^2-4\alpha\omega}]\cdot(1-\omega+\sqrt{(1-\omega)^2+4\alpha\omega})\\
&= \alpha\beta(1-\omega)(1-\omega+2\omega^2)+\alpha\beta(1-(3-2\beta)\omega+2\omega^2)(1-(1-2\alpha)\omega+2\alpha\beta\omega^2)\\
&-\alpha\beta(1+(3-2\beta)\omega+2\omega^2)(1+(1-2\alpha)\omega+2\alpha\beta\omega^2)\\
&-(1-2\omega)(1-(1-2\alpha)\omega+2\alpha\beta\omega^2)\cdot(1+(1-2\alpha)\omega+2\alpha\beta\omega^2)+O(\omega^3)\\
&=8\alpha\beta^3\omega^3+O(\omega^3).
\end{split}
\end{equation}
Hence,
\begin{equation}
AB'+C'D =
\frac{8\alpha\beta^3\omega^3+O(\omega^3)}{4\beta^2(1-2\beta(1-4\alpha)\omega^2)
+ O(\omega^3)}= 2\alpha\beta\omega^2 + O(\omega^3).
\end{equation}
Since $A'B+CD' = (AB'+C'D)'$,
\begin{equation}
A'B+CD'=2\alpha\beta\omega^2 + O(\omega^3).
\end{equation}

Hence,
\begin{equation}
\begin{split}
\inner{\varphi(T)}{\varphi'(T)} &=
(1-3\alpha\beta\omega^2)e^{i\mu_1}-\alpha\beta\omega^2e^{-i\mu_1}\\
&+ 4\alpha\beta\omega^2\cos(\mu_2) + O(\omega^3).
\end{split}
\end{equation}
So
\begin{equation}\label{inequ0}
p_s = \frac{1+|\inner{\varphi(T)}{\varphi'(T)}|^2}{2} \geq 1 -
8\alpha\beta\omega^2,
\end{equation}
and
\begin{equation}\label{inequ1}
|arg(\inner{\varphi(T)}{\varphi'(T)}) - \mu_1| \leq
8\alpha\beta\omega^2.
\end{equation}

\begin{equation}
T = O((\frac{1}{\epsilon})^{3/2})
\end{equation}

\end{document}